\documentclass[%
reprint,
superscriptaddress,
 amsmath,amssymb,
prb,
floatfix,
]{revtex4-2}
\usepackage[pdftex]{graphicx}
\usepackage{dcolumn}
\usepackage{upgreek}
\usepackage{bm}
\usepackage[colorlinks=true,linkcolor=blue]{hyperref}
\usepackage[utf8]{inputenc}
\usepackage[english]{babel}
\usepackage{xcolor}

\newcommand{\un}[1]{\,\mathrm{#1}}

\begin{document}

\title{Direct imaging of shock wave splitting in diamond at Mbar pressures}

\author{S.~S.~Makarov}
\affiliation{Joint Institute for High Temperatures of Russian Academy of Sciences, 13/2 Izhorskaya st., 125412 Moscow, Russia}
\author{S.~A.~Dyachkov}
\affiliation{Joint Institute for High Temperatures of Russian Academy of Sciences, 13/2 Izhorskaya st., 125412 Moscow, Russia}
\author{T.~A.~Pikuz}
\affiliation{Institute for Open and Transdisciplinary Research Initiative, Osaka University, Suita, Osaka 565-0871, Japan}
\author{K.~Katagiri}
\affiliation{Graduate School of Engineering, Osaka University, Suita, Osaka 565-0817, Japan}
\author{V.~V.~Zhakhovsky}
\affiliation{Joint Institute for High Temperatures of Russian Academy of Sciences, 13/2 Izhorskaya st., 125412 Moscow, Russia}
\author{N.~A.~Inogamov}
\affiliation{Landau Institute for Theoretical Physics of Russian Academy of Sciences, 1-A Akademika Semenova av., Chernogolovka, Moscow Region, 142432, Russia}
\author{V.~A.~Khokhlov}
\affiliation{Landau Institute for Theoretical Physics of Russian Academy of Sciences, 1-A Akademika Semenova av., Chernogolovka, Moscow Region, 142432, Russia}
\author{A.~S.~Martynenko}
\affiliation{Joint Institute for High Temperatures of Russian Academy of Sciences, 13/2 Izhorskaya st., 125412 Moscow, Russia}
\affiliation{Plasma Physics Department, GSI Helmholtzzentrum für Schwerionenforschung, 64291 Darmstadt, Germany}
\author{B.~Albertazzi}
\affiliation{LULI, CNRS, CEA, École Polytechnique, UPMC, Univ Paris 06: Sorbonne Universités, Institut Polytechnique de Paris, F-91128 Palaiseau cedex, France}
\author{G.~Rigon}
\affiliation{LULI, CNRS, CEA, École Polytechnique, UPMC, Univ Paris 06: Sorbonne Universités, Institut Polytechnique de Paris, F-91128 Palaiseau cedex, France}
\affiliation{Nagoya Univ, Grad Sch Sci, Chikusa Ku, Nagoya, Aichi 4648602, Japan}
\author{P.~Mabey}
\affiliation{LULI, CNRS, CEA, École Polytechnique, UPMC, Univ Paris 06: Sorbonne Universités, Institut Polytechnique de Paris, F-91128 Palaiseau cedex, France}
\affiliation{Freie Universität Berlin, Department of Physics, Experimental Biophysics and Space Sciences, Arnimallee 14, 14195 Berlin, Germany}
\author{N.~Hartley}
\affiliation{SLAC National Accelerator Laboratory, 2575 Sand Hill Road, Menlo Park, CA 94025, USA}
\author{Y. Inubushi}
\affiliation{Japan Synchrotron Radiation Research Institute, Sayo, Hyogo 679-5198, Japan}
\affiliation{RIKEN SPring-8 Center, Sayo, Hyogo 679-5148, Japan}
\author{K. Miyanishi}
\affiliation{RIKEN SPring-8 Center, Sayo, Hyogo 679-5148, Japan}
\author{K. Sueda}
\affiliation{RIKEN SPring-8 Center, Sayo, Hyogo 679-5148, Japan}
\author{T. Togashi}
\affiliation{Japan Synchrotron Radiation Research Institute, Sayo, Hyogo 679-5198, Japan}
\affiliation{RIKEN SPring-8 Center, Sayo, Hyogo 679-5148, Japan}
\author{M. Yabashi}
\affiliation{Japan Synchrotron Radiation Research Institute, Sayo, Hyogo 679-5198, Japan}
\affiliation{RIKEN SPring-8 Center, Sayo, Hyogo 679-5148, Japan}
\author{T. Yabuuchi}
\affiliation{Japan Synchrotron Radiation Research Institute, Sayo, Hyogo 679-5198, Japan}
\affiliation{RIKEN SPring-8 Center, Sayo, Hyogo 679-5148, Japan}
\author{R.~Kodama}
\affiliation{Graduate School of Engineering, Osaka University, Suita, Osaka 565-0817, Japan}
\affiliation{Institute of Laser Engineering, Osaka University, Suita, Osaka 565-0871, Japan}
\author{S.~A.~Pikuz}
\affiliation{Joint Institute for High Temperatures of Russian Academy of Sciences, 13/2 Izhorskaya st., 125412 Moscow, Russia}
\author{M.~Koenig}
\affiliation{LULI, CNRS, CEA, École Polytechnique, UPMC, Univ Paris 06: Sorbonne Universités, Institut Polytechnique de Paris, F-91128 Palaiseau cedex, France}
\affiliation{Graduate School of Engineering, Osaka University, Suita, Osaka 565-0817, Japan}
\author{N.~Ozaki}
\affiliation{Graduate School of Engineering, Osaka University, Suita, Osaka 565-0817, Japan}
\affiliation{Institute of Laser Engineering, Osaka University, Suita, Osaka 565-0871, Japan}


\begin{abstract}
The propagation of a shock wave in solids can stress them to ultra-high pressures of millions of atmospheres. Understanding the behavior of matter at these extreme pressures is essential to describe a wide range of physical phenomena, including the formation of planets, young stars and cores of super-Earths, as well as the behavior of advanced ceramic materials subjected to such stresses. Under megabar (Mbar) pressure, even a solid with high strength exhibits plastic properties, causing the shock wave to split in two. This phenomenon is described by theoretical models, but without direct experimental measurements to confirm them, their validity is still in doubt. Here, we present the results of an experiment in which the evolution of the coupled elastic-plastic wave structure in diamond was directly observed and studied with submicron spatial resolution, using the unique capabilities of the X-ray free-electron laser. The direct measurements allowed, for the first time, the fitting and validation of a strength model for diamond in the range of several Mbar by performing continuum mechanics simulations in 2D geometry. The presented experimental approach to the study of shock waves in solids opens up new possibilities for the direct verification and construction of the equations of state of matter in the ultra-high pressure range, which are relevant for the solution of a variety of problems in high energy density physics.
\end{abstract}

\maketitle

\section{Introduction}

Dynamic compression and shock loading of solid materials is a unique tool for the experimental study of the material response to ultrahigh strain rates and pressures. A transition from an elastic to inelastic response of the substance is usual at high loading pressures, when the Hugoniot elastic limit (HEL) is exceeded. At this stage, the shock wave may split into an elastic precursor and a slower plastic shock wave that follows it~\cite{2020:PRB:Winey,2004:book:Kanel}. The study of elastic-plastic shock waves is of great importance in solid mechanics being the main source of data for dynamic material strength models~\cite{1980:LASL.shock}. The latter may be used to analyze a wide range of phenomena from engineering applications to the formation of planets from aggregation of small objects (such as meteorites) at high velocity collisions.

To track directly shock wave evolution, in particular the elastic-plastic wave splitiing in solids, it is necessary to have a transparent material. At the same time a material should withstand ultra-high strain rates and pressures. Diamond perfectly corresponds to these requirements. Its unique combination of ultra-high stiffness, hardness, optical transparency and thermal conductivity makes it a popular research object when high pressures in matter are considered~\cite{2008:PRB:Hicks,2008:Science:Knudson,2010:PRB:McWilliams,2010:NaturePhys:Eggert,2014:Nature:Smith}. However, in an extreme environment exceeding several Mbar, its physical characteristics are not well known even though diamond is used for high-pressure anvil cells (DACs) and the like. The matter dynamically compressed to Mbar pressures moves at a speed approaching several tens of km/s, and it is necessary to observe the stress state of the lattice and its temporal change in real time. A broad and deep understanding of ``diamond in the extreme environment'' up to off-Hugoniot is required for improving DAC performance, the internal structure of giant planets, warm dense matter (WDM) characteristics, and the behavior of laser fusion fuel shells. 

Commonly used at high pressures physics methods such as the Velocity Interferometer System for Any Reflector (VISAR)~\cite{2020:PRB:Winey,2000:AIPConfProc:Barker,2019:MSRU:Ziborov} and Photon Doppler Velocimetry (PDV)~\cite{2020:PRB:Winey,2006:RevSciInstr:Strand,2011:RevSciInstr:Ao} do not allow direct measurements but give only a limited idea of how shock waves behave inside a sample. Therefore, the state of matter as the shock wave propagates inside can only be assessed by an indirect method. Absorption and phase-contrast X-ray methods based on a laser-plasma source~\cite{2018:JInstr:Antonelli,2019:SciReports:Barbato} give a low image contrast, which is not enough to clearly resolve the regions inside the shock wave or the plastic shock wave. Therefore, it is difficult to establish accurate equations of state and verify theoretical models for the response of a substance under ultrahigh pressure conditions. 

The unique parameters of pulses of the X-ray free electron lasers (XFELs) opened a new branch in the study of matter under ultrahigh pressure. In particular, femtosecond XFEL pulses began to be used in X-ray diffraction method (XRD) for measurement of shock-wave-driven twinning and lattice dynamics of tantalum~\cite{2017:Nature:Wehrenberg} and the dynamic fracture of tantalum under extreme tensile stress~\cite{2017:SciAdv:Albertazzi}. By combining focused XFEL beam and a high-power laser, X-ray phase contrast imaging (PCI) observation is already possible even in the dynamic ultra-high pressure extreme environment in the sub-TPa region~\cite{2015:SciRep:Schropp}.

\begin{figure}[t]
\includegraphics[width=0.98\linewidth]{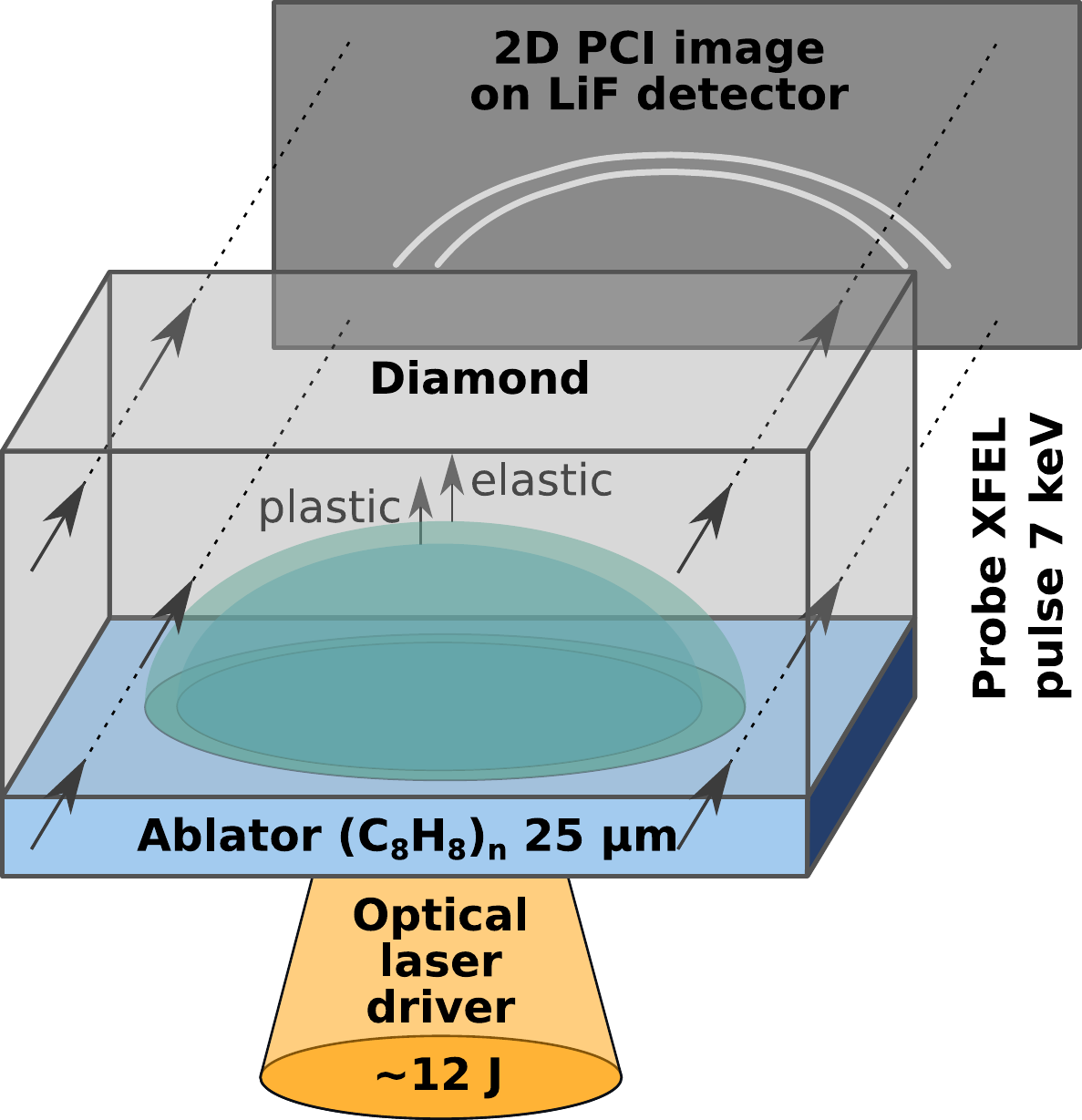}
\caption {\label{fig:setup} Outline of pump-probe experiment for visualization of elastic-plastic shock waves evolution in diamond with submicron spatial resolution. A shock wave is driven by a focused pump laser (red arrow) into the target consisting of an ablator (25-$\mu$m-thick polystyrene) and 200-$\mu$m-thick monocrystalline diamond with crystallographic orientation $<$100$>$ along the propagation direction of the laser. Hard X-ray beam (XFEL, green) probes the target with nanosecond time delay with respect to the driver laser to observe the dynamic of shock wave into diamond. To resolve the morphology of the low-contrast elastic-plastic shock waves with submicron spatial resolution, a LiF detector was used. For observing Phase-Contrast-enhanced radiography images, a detector was put at the distance $\sim 110\un{mm}$ from the diamond target (along the XFEL beam propagation direction).}
\end{figure}

Here we use an X-ray free-electron laser source to make comparisons between direct observation in experiment and hydrodynamic simulations of wave splitting into an elastic shock precursor and a plastic shock wave in diamond. The X-ray platform with a parallel XFEL beam and a lithium fluorine (LiF) fluorescent detector, which was developed in Ref.~\cite{2018:SciRep:Faenov} and has been successfully used in recent studies of microscale phenomena in plasma~\cite{2021:NatComm:Rigon}. The experiment was performed at SPring-8 Angstrom Compact Free Electron Laser (SACLA XFEL, Japan) at experimental hutch EH5. A sketch of the experimental setup is shown in Fig.~\ref{fig:setup}. The nanosecond optical driver pulse ($\sim12\un{J}$, $5\un{ns}$, FWHM at sample $250\un{\mu m}$) focused on a multilayer target containing a thin layer of polystyrene ablator under the diamond sample and loaded it up to several Mbar pressures. Such a target geometry makes it possible to achieve the best conditions for generating elastic-plastic shock waves in diamond. The evolution of the shock waves in the target was temporally resolved by irradiating the XFEL pulse (photon energy $7\un{keV}$, pulse duration $8\un{fs}$, FWHM in the target region $600\un{\mu m}$) with changing the delay relative to the optical drive laser irradiation timing.The femtosecond duration of X-ray pulse allowed to probe sample before structural changes could occur in the lattice of a diamond. A fluorescent crystal detector LiF was used to register the formed phase-contrast pattern with a submicron spatial resolution. 

To confirm the observed emerging of the elastic-plastic structure in diamond we performed continuum mechanics modelling. The proper laser pump interaction with an ablator is obtained using 1D MULTI code~\cite{1988:CPC:Ramis}. The parameters of the emerging shock wave at the ablator-diamond interface are approximated to be used as a boundary condition in continuum mechanics with strength. The SPH method~\cite{2002:JCP:Parshikov} and our in-house code~\cite{2019:CPC:Egorova} were used to study the shock propagation in the diamond bulk in accordance with the experimental setup.

\section{The experiment: setup, diagnostics, and results}

\subsection{Target composition details}

The multi-layer target is used for our experiments composed of a polystyrene ($1\un{g/cm^3}$) ablator and a monocrystalline diamond sample ($3.51\un{g/cm^3}$) with dimensions of $\Delta x \times \Delta y \times \Delta z = 2000 \un{\mu m} \times 25\un{\mu m}\times 2000 \un{\mu m}$ and $\Delta x \times \Delta y \times \Delta z = 1500\un{\mu m} \times 210\un{\mu m}\times 1500\un{\mu m}$ respectively. Diamond crystallographic orientation is $<$100$>$ along the shock direction (it corresponds also to the direction of a driver optical laser) and $<$010$>$ along the XFEL irradiation direction. Diamond samples were made by chemical vapor deposition and both $1500\un{\mu m} \times 1500 \un{\mu m}$ surfaces were polished.

\subsection{Initiation of elastic-plastic shock waves in target}

\begin{figure}
\includegraphics[width=0.909\linewidth]{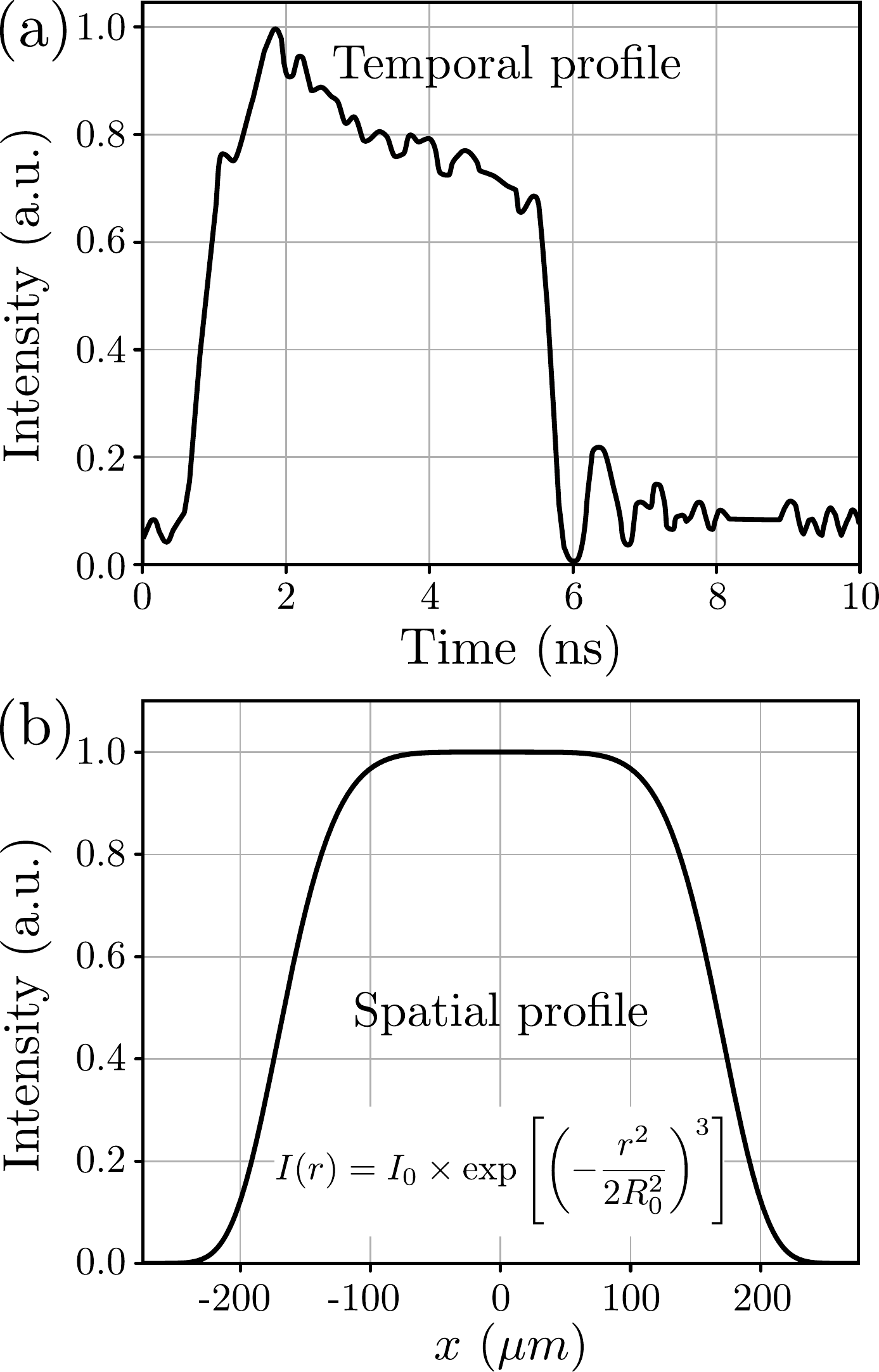}
\caption {\label{fig:laser} Driver laser intensity distribution $I(x,y,t)$: (a) Temporal distribution of optical laser. (b) Spatial distribution of a focused laser at the plane of the polystyrene ablator.}
\end{figure}

Focused optical laser with a wavelength $\lambda = 510\un{nm}$, a square pulse with duration of $t = 5\un{ns}$, energy $E = 9-12\un{J/pulse}$ and Super Flat-Gaussian-shape:
\begin{equation}
\label{eq:intensity-profile}
I(r) = I_0 \times \exp \left[ \left(-\frac{r^2}{2 R_0^2}\right)^3 \right]
\end{equation}
where $r$ is the distance from its center, $I_0$ is the peak intensity, $R_0 = 125\un{\mu m}$) was used for initiating shock waves in target. The temporal and spatial profiles of optical lasers are shown in Fig.~\ref{fig:laser}. The drive laser was focused on the ablator with a spot size of $250 \un{\mu m}$, corresponding to a maximum peak power of $P = 6-9\un{TW/cm^2}$ and a maximum pressure of up to $2-4\un{MBar}$ (see MULTI simulations) in the polystyrene ablator.

\subsection{Compact Phase-Contrast Imaging by using LiF detector and XFEL}

\begin{figure}[t]
\includegraphics[width=0.98\linewidth]{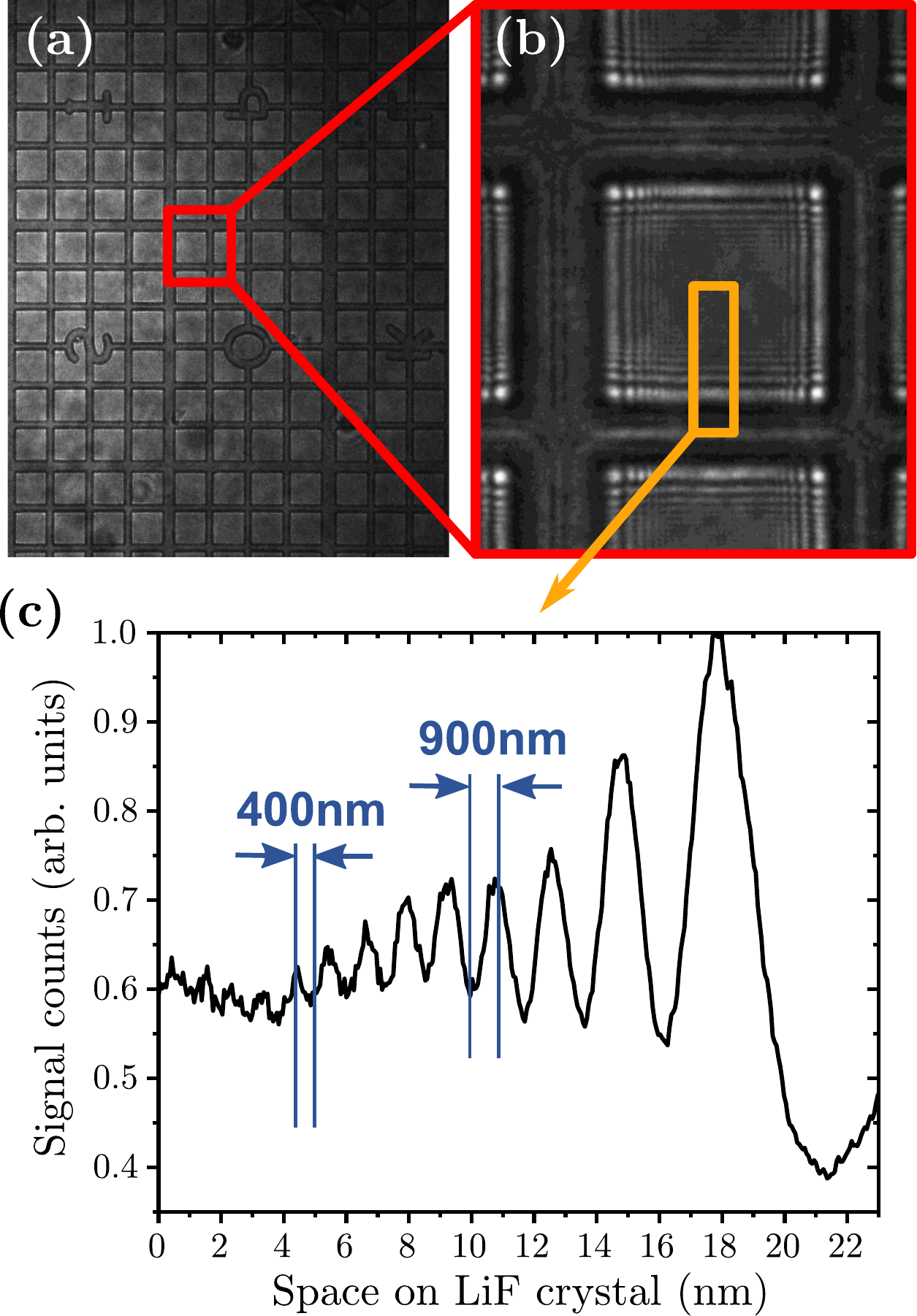}
\caption {\label{fig:LiF-PCI-platform} The spatial resolution of the PCI platform for experimental geometry used in the main article: X-ray image of the Ni mesh ($\mathrm{lpi} = 400$) irradiated with direct XFEL SACLA beam (photon energy $7\un{keV}$) and obtained on the LiF detector at a distance of $110\un{mm}$ with different magnification: (a) $40\times$ and (b) $100\times$; (c) profile of the experimental diffraction pattern along the red line in (b). These images show that the spatial resolution in the plane of the LiF detector is several hundred nanometres.}
\end{figure}

The phase contrast effect can significantly increase the information content of X-ray images of low-contrast objects that cannot be resolved with conventional absorption radiography. The appearance of phase contrast in an image is determined by the phase shift of the X-ray photons as they pass through objects with a strong density gradient. Under optimal conditions, this appears on the image as characteristic black and white diffraction fringes. 

A compact PCI platform with submicron resolution was used to visualise laser-induced shock waves in the target. This platform was developed in Ref.~\cite{2018:SciRep:Faenov} and used both the SACLA XFEL beam and a fluorescent crystal detector: lithium fluorine (LiF). In a recent study~\cite{2021:NatComm:Rigon} it was also used for the first time to visualize the evolution of the Rayleigh--Taylor instability in a laser-induced plasma to a turbulent phase with a micrometer scale and a stage of energy dissipation. For our purposes, we used a collimated SACLA XFEL beam with a divergence angle of $\sim 2\un{\mu rad}$ and a Gaussian intensity profile ($\mathrm{FWHM} = 600\un{\mu m}$ in the plane of the diamond sample) and a photon energy of $E = 7\un{keV}$. An ultra-short pulse duration of the XFEL beam provides a temporal resolution of the PCI platform in the femtosecond range (it corresponds to the pulse duration of the probe beam $t \sim 8\un{fs}$). The LiF crystal was placed $\sim 110\un{mm}$ after the diamond sample, giving an optimal spatial resolution of $0.4\un{\mu m}$ in our experimental geometry (see Fig.~\ref{fig:LiF-PCI-platform}).

As an example for our experimental conditions, if to make an estimation of XFEL intensity passing through the diamond sample in for two simple cases for photon energy $7\un{keV}$: 
\begin{enumerate}
\item Unshocked diamond $1500\un{\mu m}$ (target thickness along the XFEL probe direction), density $3.51\un{g/cm^2}$: transmission $T = 3.1$\%. 
\item Unshocked diamond $\sim$~$1250\un{\mu m}$, density $3.51\un{g/cm^2}$ + excited diamond $\sim 250\un{\mu m}$, density $4.3\un{g/cm^2}$: $T = 2.6\%$ (real shocked volume is much thinner). 
\end{enumerate}

Thus, the absorption contrast is very small ($\Delta T < 0.5\%$) and the shock waves can be visible only due to a phase contrast effect in our experiment.

\subsection{Observation of the shock propagation}

\begin{figure}[t]
\includegraphics[width=0.98\linewidth]{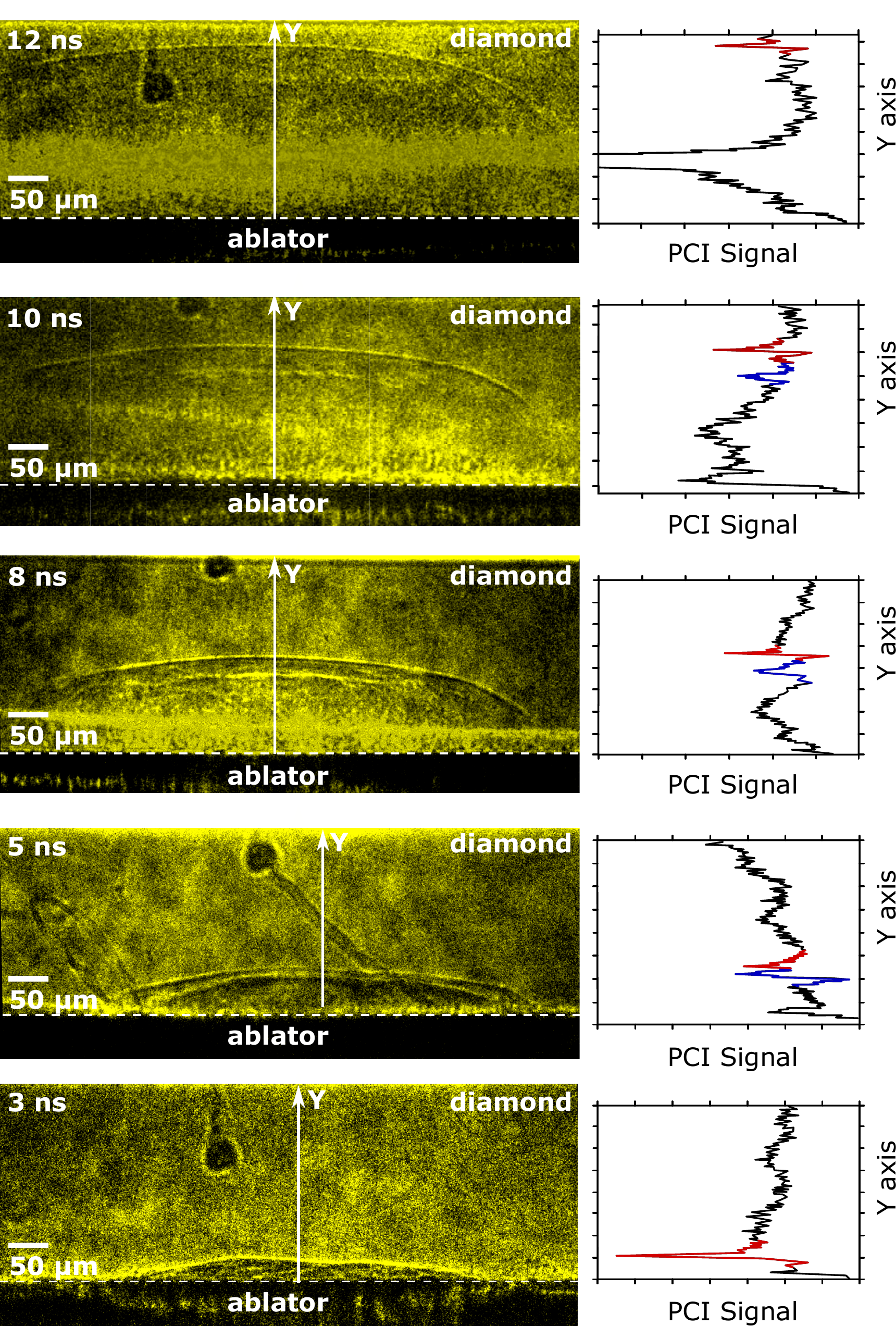}
\caption {\label{fig:wave-propagation-exp} Dynamics of shock wave evolution for times $t = 3$--$12\un{ns}$ after interaction of an optical laser on the target: Phase contrast images of SW evolution in diamond taken with a LiF detector located at a distance of $110\un{mm}$ from the target. Corresponding PCI signal intensity profiles taken along the Y direction showing the manifestation of phase contrast at the impact fronts.}
\end{figure}

\begin{figure}[t]
\includegraphics[width=1\linewidth]{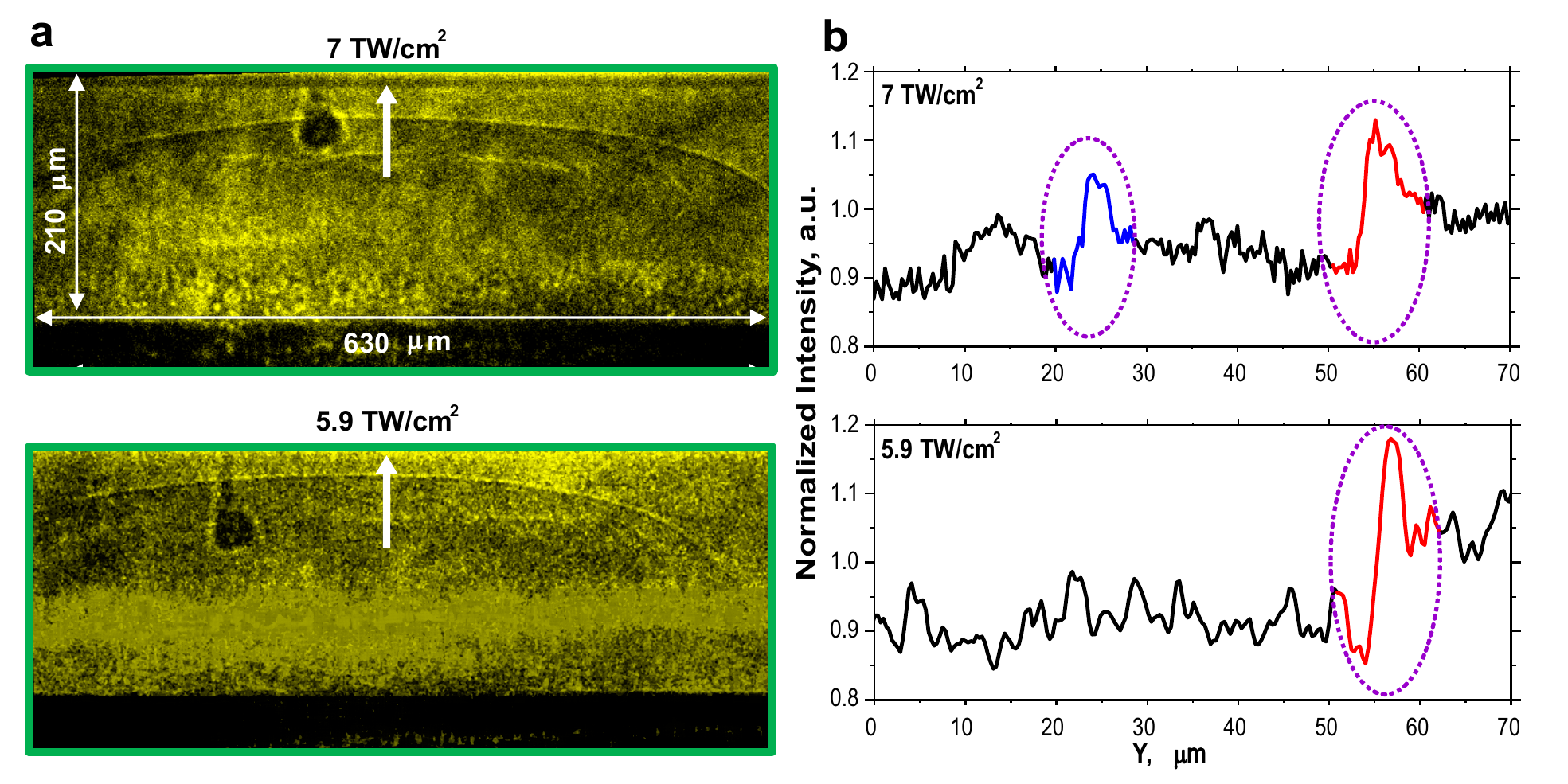}
\caption {\label{fig:12ns_diff_intens} The structure of the shock wave at time $t = 12\un{ns}$ after the interaction of an optical laser with an intensity of I = $7\un{TW/cm^2}$ and $5.9\un{TW/cm^2}$ on the target: (a) Phase contrast images recorded on LiF. (b) The intensity of the PCI signal, traced along the Y-direction into red rectangle in case (a).}
\end{figure}

The evolution of shock waves in the diamond sample was traced up to $12\un{ns}$ after the beginning of the main laser pulse. In Fig.~\ref{fig:wave-propagation-exp}(left) we present a series of phase-contrast images recorded by using a LiF detector at different delay times in the range from $3\un{ns}$ to $12\un{ns}$. The phase-contrast enhancement and the submicron resolution of our experimental platform allowed to clearly resolve the front of the generated shock waves (the difference in absorption in a shocked and unshocked diamond area is less than 0.5\%). In the region behind the plastic wave, the remaining trace of plastic deformations is visible. Such an observation is unprecedented and represents a new horizon in the model development and validation of pressure-driven shock wave simulations.

At the initial stage of waves development  ($t = 3\un{ns}$) one can observe only a single shock wave in Fig.~\ref{fig:wave-propagation-exp} which is probably an elastic-plastic one without notable wave splitting. At times between $3\un{ns} < t < 5\un{ns}$, the shock wave splits into a clear two-wave structure in the diamond bulk due to the difference in elastic and plastic wave speeds~\cite{2004:book:Kanel}. Such a regime emerges when $P_\mathrm{HEL}$ is exceeded: a plastic wave appears and begins to propagate in the elastically compressed material with the bulk sound speed according to the equation of state, while the elastic precursor outruns such wave being enforced by shear stresses. 

The observed shock wave is supported by the laser pulse for several nanoseconds, after which the release wave propagating from the ablator side reduces the plastic wave amplitude. As a result, the plastic wave front disappears completely between $10\un{ns} < t < 12\un{ns}$ as shown in Fig.~\ref{fig:wave-propagation-exp}. The increase of the pump laser intensity to~$7\un{TW/cm^2}$ leads to the increase in shock wave amplitude, so that the plastic wave can still be observed at time $t = 12\un{ns}$ (see Fig.~\ref{fig:12ns_diff_intens}). It is also seen on the intensity profile obtained from the LiF image (see Fig.~\ref{fig:12ns_diff_intens}b): the plastic shock wave appears at time $t = 12\un{ns}$ with the laser intensity change from $5.9\un{TW/cm^2}$ to $7\un{TW/cm^2}$.

Using the data on the position of the shock wave fronts obtained from the radiographic LiF images, the velocities of the SWs observed in the experiment were reconstructed as they propagate inside the diamond, Fig.~\ref{fig:piston-MULTI-vs-SPH}c. In Fig.~\ref{fig:piston-MULTI-vs-SPH}c, red and blue markers indicate the obtained velocities for elastic and plastic shock waves, respectively. It can be seen that the velocity of the precursor does not change as it passes through the diamond ($V_\mathrm{elastic} = 19\pm0.5\un{km/s}$), while the plastic SW slows down (from $V_\mathrm{plastic} = 17.2\pm0.5\un{km/s}$ to $15.2\pm0.5\un{km/s}$) and disappears at times about $10\un{ns}$.

\section{Simulation of the laser pulse absorption}

\subsection{MULTI simulations}

Modeling of laser ablation and shock wave generation in polystyrene is performed using a one-dimensional radiation hydrodynamics code MULTI~\cite{1988:CPC:Ramis}. For the simulations, the SESAME table No. 7590 is used for polystyrene ablator (its gross chemical formula is $(C_{8}H_{8})_n$) and SESAME table No. 7830 for diamond with the initial densities of $\rho_1 = 1.1\un{g/cm^3}$ and $\rho_2 = 3.52\un{g/cm^3}$, respectively. The ablator thickness is set to $25\mu m$.

\begin{figure}[t]
\includegraphics[width=0.98\linewidth]{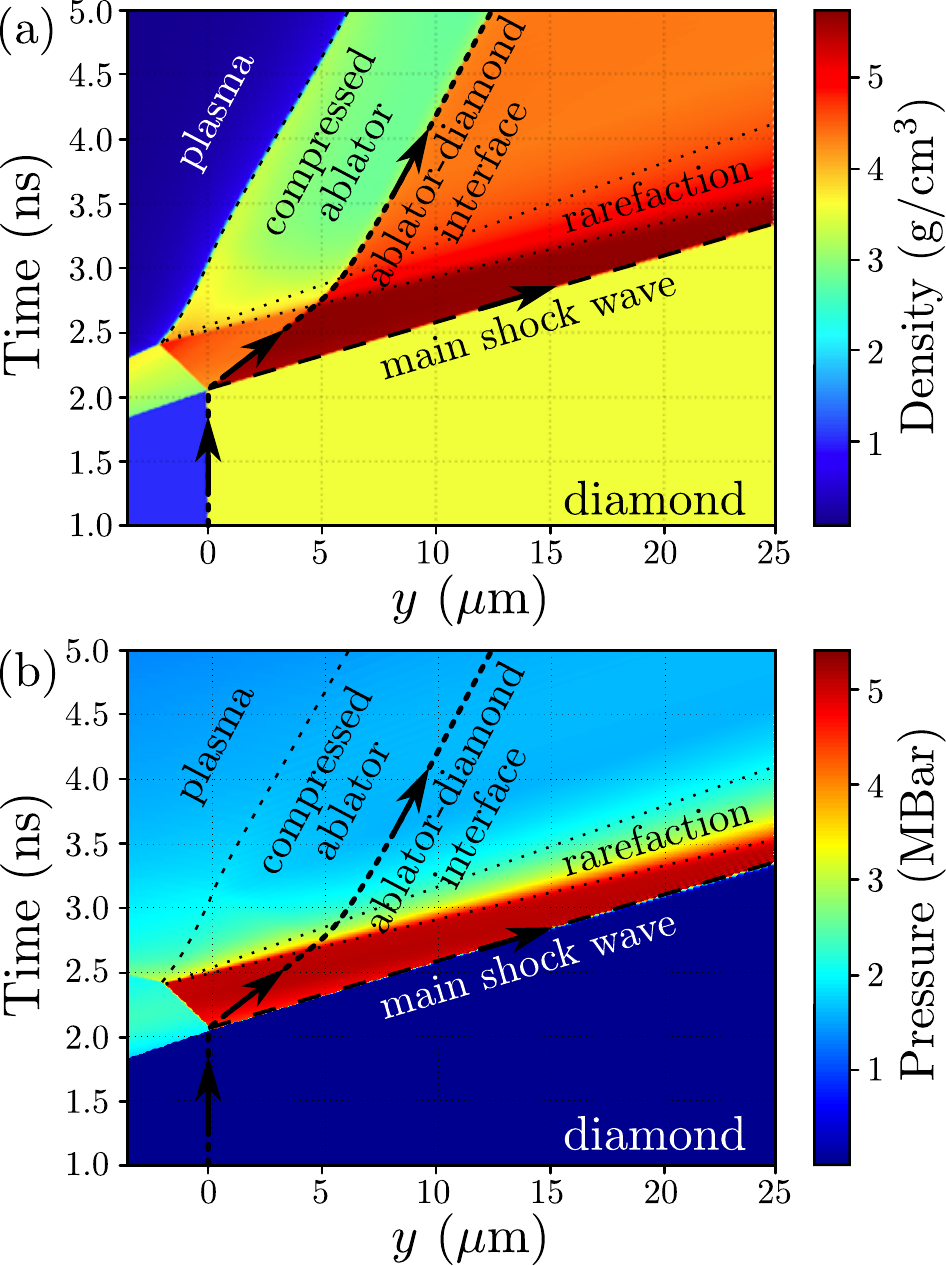}
\caption {\label{fig:MULTI} (a) Density and (b) pressure maps obtained in the one-dimensional simulation using radiation hydrodynamics code MULTI.}
\end{figure}

Figure~\ref{fig:MULTI} shows the $xt$-diagrams as colormaps for the density and the pressure from a particular MULTI simulation. Shown are the hydrodynamic processes occurring in the target at time and space intervals of $1$--$5\un{ns}$ and $20$--$50\un{\mu m}$, respectively. The position $0\un{\mu m}$ corresponds to the ablator-diamond interface, while the ``front'' part of the ablator at which the laser pulse arrives at time $0\un{ns}$ (~1\% of the maximum laser intensity) is placed at $x = -25\un{\mu m}$.

The laser pulse (LP) is absorbed in the polystyrene ablator resulting in extreme heating and pressure growth. The ablated surface is evaporated and ionized producing plasma, so that LP continues to be absorbed in the area of the corona with a critical density (about 2 orders of magnitude lower than the value in solid ablator). LP of the intensity $I = 10^{13}\un{W/cm^2}$ produces the pressure in the corona about $2\un{Mbar}$ which keeps the bulk of the ablator from unloading until the end of LP. An initial shock wave propagates along the ablator to the interface with diamond: the ablator layer is compressed by $\sim 3$--$3.5$ times (its thikness changes from $25\mu m$ to $\sim 6$--$8\un{\mu m}$) leading to the pressure growth up to $\sim 2\un{Mbar}$, which is close to the pressure at the laser ablation front as shown in Fig.~\ref{fig:MULTI} at $2$--$2.5\un{ns}$.

\begin{figure*}[t]
\includegraphics[width=1\textwidth]{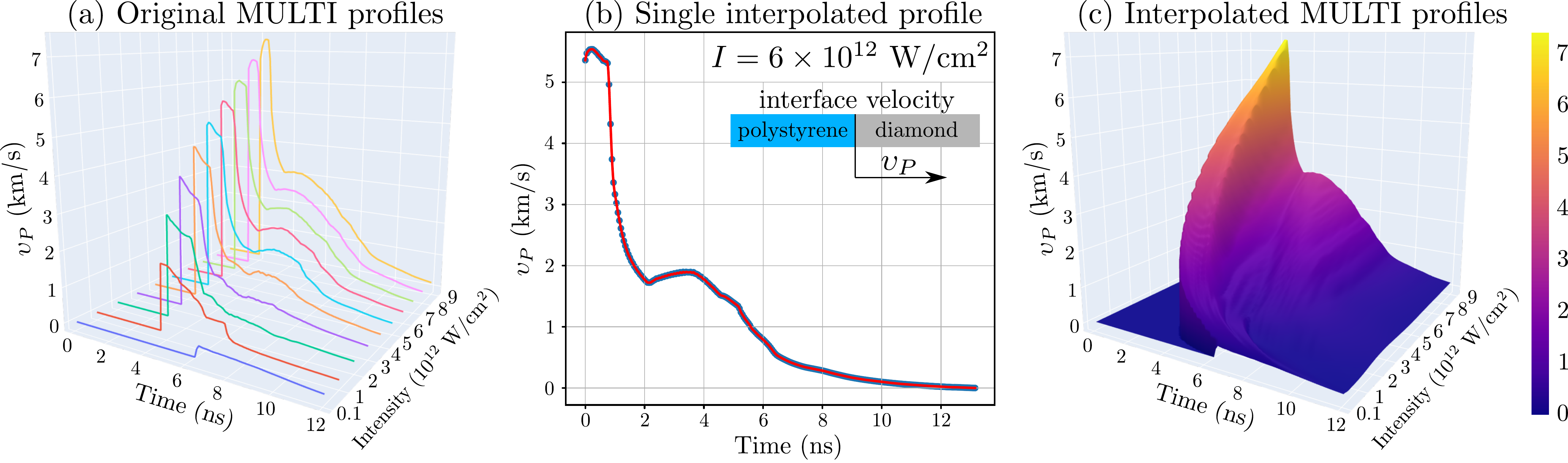}
\caption {\label{fig:multi-interp} (a) The velocity profiles $\upsilon_P(t)$ of the ablator-diamond interface obtained in 1D MULTI simulations for various intensities. (b) The result of the velocity profiles interpolation $\upsilon_P(t, I)$ required for further multi-dimensional SPH simulations with strength.}
\end{figure*}

The shock wave reaches the interface between polystyrene and diamond at $\sim 2.05\un{ns}$. The main shock wave begins to propagate along the diamond sample, while the reflected shock wave begins to propagate through the precompressed polystyrene moving in the opposite direction from the diamond boundary to the ablation front. This reflected shock wave passes the thin layer of the shocked ablator in $0.4\un{ns}$ and reaches the ablation front. Afterwards, the SW is reflected by this boundary and produces the spray of rarefaction waves resulting in the ablator expansion. The rarefaction wave reaches the surface of the diamond at $\sim 2.7\un{ns}$ and follows the main shock wave in the diamond bulk.

One can note the ``triangle'' of high density in the ablator formed by the aforementioned shock wave which is reflected from the ablator-diamond interface. The density in this triangle exceeds the initial density by $4.5$--$5$ times producing the pressure about $\sim 4\un{Mbar}$ which is about 2 times higher than the pressure in the plasma at the ablation front.

The release wave reaches the interface at $\sim 2.7\un{ns}$ which is less than the pulse duration ($5\un{ns}$). This leads to the pressure drop from $\sim 4\un{Mbar}$ to the pressure of the laser corona ($\sim 2\un{Mbar}$) which sustains until the end of the laser pulse duration. The velocity of the ablator-diamond interface moves according to the applied pump: it accelerates up to $\sim 7\un{km/s}$ by $2.7\un{ns}$. The end of the laser pulse is followed by the gradual decrease of pressure in the ablated plasma. As a result, the unburnt part of the ablator begins to release and is pushed from the interface. The pressure on the diamond surface remains for a few tenths of nanoseconds until the ``signal'' about the end of the laser heating of the corona and zero pressure at the edge of the unburnt ablator reaches the diamond. The release leads to a gradual decrease of the interface velocity to almost $0\un{km/s}$.

The above mechanism is realized for laser pulse intensities starting from $10^{12}\un{W/cm^2}$ and higher. For lower intensities of the order $10^{11}\un{W/cm^2}$, the shock wave passing through the ablator reaches the ablator-diamond interface only after the end of the laser pulse.

The pressure $P$ of the ablated plasma can be estimated using a well-known scaling-law~\cite{1995:PhysPlasmas:Lindl}:
\begin{equation}
    P = 8.6 \times \left(\frac{I}{10^{14}}\right)^{2/3} \lambda^{-2/3}\left(\frac{A}{2Z}\right)^{1/3},
\end{equation}
where $P$ is in Mbar, the laser intensity I is in $\un{W/cm^2}$, the laser wavelength $\lambda$ is in $\un{\mu m}$; $A$ and $Z$ are the atomic mass number and atomic number of the target material, respectively.

\subsection{The ablator response model}

As it is mentioned earlier, the pressure pulse produced by the ablator on a diamond sample is simulated using the one-dimensional MULTI code. However, in a multi-dimensional (2D or 3D) case one has to take into account the spatial distribution of the ablator-diamond interface velocity $\upsilon_P$. The laser intensity in our experiments is supposed to have the super-gaussian distribution~\eqref{eq:intensity-profile}. To reproduce an adequate response of a multi-dimensional ablator we model the interface velocity $\upsilon_P(t)$ at various laser intensities $I \in (0, I_0]$ in one-dimensional MULTI code, which are interpolated for $\upsilon_P(t, I)$. The latter function is then transformed to $\upsilon_P(r, t) = \upsilon_P(t, I(r))$ and can be applied to model a multi-dimensional boundary condition at the ablator-diamond interface.

\begin{figure*}[t]
\includegraphics[width=1\textwidth]{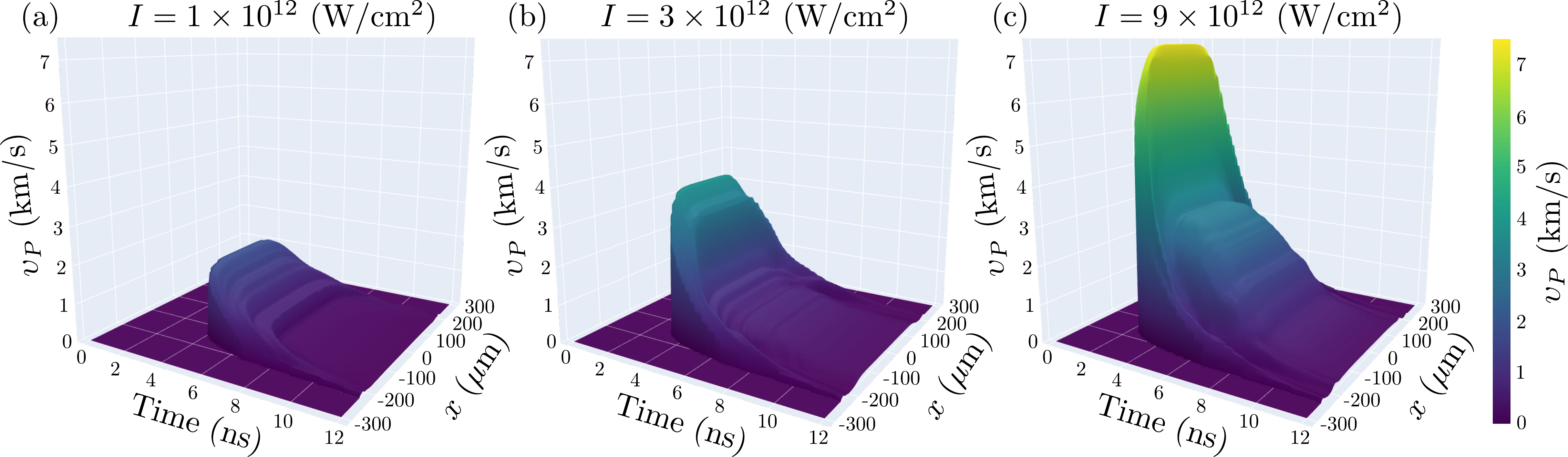}
\caption {\label{fig:multi-piston-intensity} The velocity profiles $\upsilon_P(x, t)$ ($x$ is the axis directed along the spot diameter) obtained according to the spatial laser intensity profile~\eqref{eq:intensity-profile} for different peak intensities $I_0$: (a) $I_0 = 1 \times 10^{12}\un{W/cm^2}$; (b) $I_0 = 3 \times 10^{12}\un{W/cm^2}$; (c) $I_0 = 9 \times 10^{12}\un{W/cm^2}$.}
\end{figure*}

Fig.~\ref{fig:multi-interp} shows the result of MULTI profiles simulation and their interpolation. To construct $\upsilon_P(t, I)$ we calculated 10 interface velocity profiles in MULTI for the intensities $I \in [0.1, 1, 2, 3, 4, 5, 6, 7, 8, 9] \times 10^{12}\un{W/cm^2}$  as shown in Fig.~\ref{fig:multi-interp}(a). As a result, the obtained interpolation $\upsilon_P(t, I)$ is shown in Figure~\ref{fig:multi-interp}(b). One may notice, that the higher intensity pulses arrive to the ablator-diamond interface faster due to a dramatic change in the ablator sound velocity at high compression. The arrival of the shock from the ablator provides an extreme growth of velocity up to several km/s, which is followed by a small plateau and a gradual release. The shape of the plateau repeats the laser intensity profile shown in Fig.~\ref{fig:laser}, but its length is reduced with the intensity growth. The release at high intensities is interrupted by the pressure growth in coronal plasma. 

The transition from the function $\upsilon_P(t, I)$ to $\upsilon_P(r, t)$ according to the distribution~\eqref{eq:intensity-profile} is presented in Fig.~\ref{fig:multi-piston-intensity} for several peak intensities $I_0$, where $x$ is the axis directed along the spot diameter. One may notice the formation of the most intense load around the center of the spot, which is followed by the fast release. However, one may notice a ring at the periphery which produces some load after the release in the center due to later arrival of a peripheric shock wave. The latter is clearly seen in Fig.~\ref{fig:multi-piston-intensity}(b). It is unclear, whether this effect is an artificial result of the interpolation, or it may appear in real simulation of 2D (3D) laser radiation absorption. Nevertheless, such small distortion at the periphery does not affect the propagation of a main shock wave in diamond as we can see from the following SPH simulation results. 

\begin{figure*}[t]
\includegraphics[width=1\textwidth]{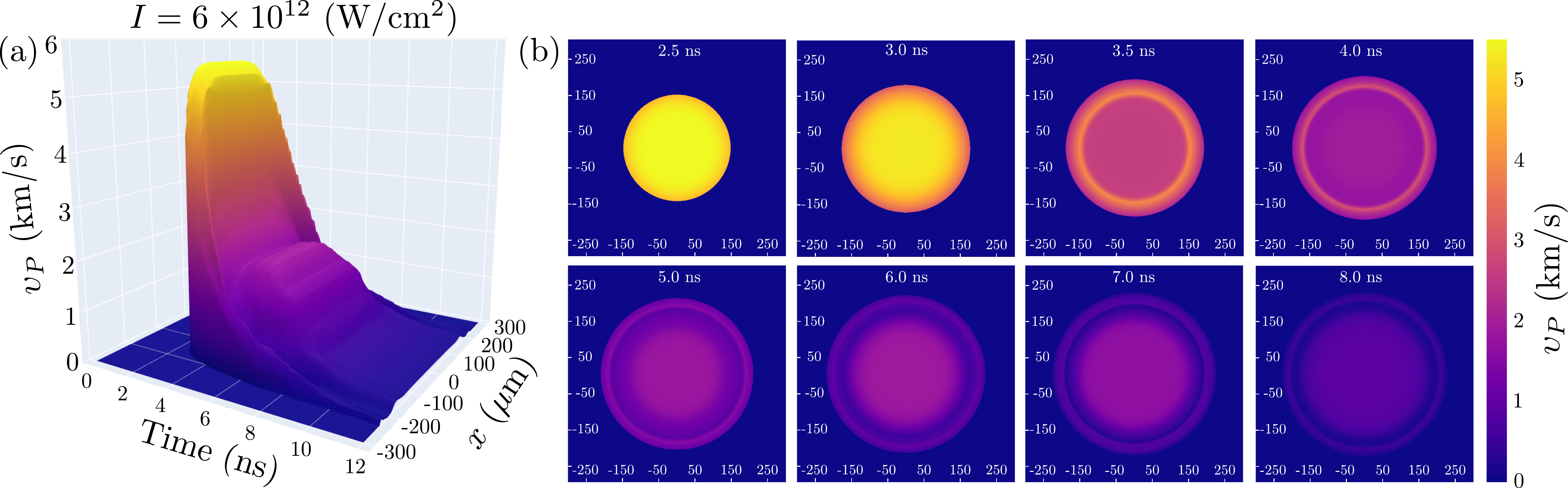}
\caption {\label{fig:multi-piston-time} (a) The velocity profile $\upsilon_P(x, t)$ ($x$ is the axis directed along the spot diameter) obtained according to the spatial laser intensity profile~\eqref{eq:intensity-profile} for $I_0 = 6 \times 10^{12}\un{W/cm^2}$. (b) The sequence of colormaps of spatial velocity distribution corresponding the velocity profile on the left. }
\end{figure*}

\section{Simulation of the shock propagation}

Smoothed particle hydrodynamics (SPH) is widely used to model compressible media with strength at extremes. In such conditions, propagation of shock waves may be accompanied by the development of instabilities, formation of cavities, material spallation and fracture, which are difficult to model using the conventional eulerian or lagrangian methods on a mesh. Most eulerian codes and the aforementioned lagrangian MULTI code are also lacking the material strength, which is necessary to model the splitting of elastic and plastic shock waves. The meshless SPH approach allows to model such phenomena naturally, without using complex algorithms for capturing the interfaces and the free boundaries, while its lagrangian formulation leads to adaptation of particle sizes in accordance with the material strain. Here we provide a brief overview of the underlying material model and SPH method used to model shock propagation in diamond induced by the ablated layer of polystyrene.

\subsection{The governing equations}

The evolution of continuous material with strength is guided by the equations which express the conservation of mass, momentum, and energy:
\begin{gather}
\label{continuity}
\frac{1}{\rho} \frac{d\rho}{dt} = -\dot\theta=-\nabla \cdot \mathbf{U}, \\
\label{momentum}
\rho \frac{d\rho}{dt} = \nabla \cdot \bm{\upsigma}, \\
\label{energy}
\rho  \frac{dE}{dt} = \nabla \cdot \left(\bm{\upsigma} \cdot \mathbf{U}\right).
\end{gather}

Here $\rho$ is the density, $\mathbf{U}$ is the velocity, $E = e + \mathbf{U}^2/2$ is the specific total energy of a material element consisting of internal and kinetic energy, and $\bm{\upsigma}$ is the total stress tensor:
\begin{equation}
\bm{\upsigma} = -P(\rho, e)\mathbf{I} + \mathbf{S},
\end{equation}
which consists of diagonal elements representing the pressure $P$ ($\mathbf{I}$ is the unity matrix) and the elastic stress deviator $\mathbf{S}$ is a symmetric tensor which trace equals to zero
\begin{equation}
S_{xx} + S_{yy} + S_{zz} = 0.    
\end{equation}
The stress deviator S is evaluated according to the Hooke’s law:
\begin{equation}
\label{eq:stress-tensor}
\dot{\mathbf{S}}=2G(\dot{\bm{\upvarepsilon}}-\dot{\bm{\upvarepsilon}} \otimes \mathbf{I}/3)- \dot{\mathbf{\Omega}}\cdot \mathbf{S}+ \mathbf{S}\cdot \dot{\mathbf{\Omega}},
\end{equation}
Here $G$ is the shear modulus, $\dot{\bm{\upvarepsilon}}$ is the strain tensor which is defined according the Saint-Venant's compatibility condition
\begin{equation}
\label{eq:strain-rate}
\dot{\bm{\upvarepsilon}} = \frac{1}{2}\left[\left( \nabla \otimes \mathbf {U} \right)^{\mathrm{T}} + \nabla \otimes \mathbf{U} \right].
\end{equation}
while the angular velocity tensor $\mathbf{\Omega}$ is given by
\begin{equation}
\label{eq:angular-velocity}
\mathbf{\Omega} = \frac{1}{2}\left[\left( \nabla \otimes \mathbf {U} \right)^{\mathrm{T}} - \nabla \otimes \mathbf{U} \right]
\end{equation}

The elastic loading is limited by the Hugoniot Elastic Limit (HEL), below which the stress growth is linearly proportional to the strain, while exceeding of it leads to the plastic strain growth. The behavior of materials subjected to plastic strain is guided by the equation of state, which expresses the relationship between $P$, $\rho$, $e$, for example, in the form of the function $P(\rho,e)$. The stress tensor $\bm{\upsigma}$ which is realized at the transition from elastic to plastic state, should satisfy the von Mises yield criterion: 
\begin{equation}
\label{eq:yield-criterion}
    \sigma_e < Y,
\end{equation}
where $Y$ is the shear (yield) strength and $\sigma_e$ is the stress tensor invariant (the equivalent stress):
\begin{multline}
  \label{sigma_equiv}
  \sigma^2_e = \frac{3}{2} S^{\alpha \beta}S^{\beta \alpha} = \frac{1}{2} \left\{(S^{xx} - S^{yy})^2 + (S^{yy} - S^{zz})^2 \right. +\\+ \left.
 (S^{xx} - S^{zz})^2 + 6\left[(S^{xy})^2 + (S^{xz})^2 + (S^{yz})^2\right]\right\},
\end{multline}
with subscripts $\alpha=x,y,z$, $\beta=x,y,z$.

In the case of uniaxial compression along x direction $\sigma_\mathrm{HEL}$ may be expressed as:
\begin{equation}
    \sigma_\mathrm{HEL} = P_\mathrm{HEL} - S_{xx}
\end{equation}
while the stress deviator components should be $S_{yy} = S_{zz} = -S_{xx}/2$. In this case, one may obtain a simple expression for $\sigma_e$:
\begin{equation}
    \sigma_e = \frac{3}{2}|S_{xx}| = \frac{3}{2}|\sigma_\mathrm{HEL} - P_\mathrm{HEL}|,
\end{equation}
which is an estimate for the material yield strength $Y$ obtained in Hugoniot measurements. The most simple strength model corresponds to $Y = \mathrm{const}$, however, the shear strength may depend on various factors, such as strain rate, temperature, etc.

Wave splitting in elastic-plastic medium appears due to difference in elastic and plastic (bulk) wave speeds~\cite{2004:book:Kanel}. Bulk sound speed $c_b$ is a property of the equation of state:
\begin{equation}
    c_b^2 = \left(\frac{\partial P}{\partial \rho}\right)_S, \quad B = \rho c_b^2,
\end{equation}
where $S$ is the entropy, $B$ is the bulk modulus. The bulk and shear moduli are usually related via the Poisson coefficient $\eta$, so that
\begin{equation}
    G = \frac{3}{2} \frac{1 - 2 \eta}{1 + \eta} B.
\end{equation}
Thus, the longitudinal $c_l$ and transversal $c_t$ wave speeds in elastic media are
\begin{equation}
    c_l^2 = \frac{1}{\rho} \left(B + \frac43 G\right) = c_b^2 + \frac{4}{3}\frac{G}{\rho} = 3 c_b^2 \frac{1 - \eta}{1 + \eta},
\end{equation}
\begin{equation}
    c_t^2 = \frac{G}{\rho} = \frac{3}{2} c_b^2 \frac{1 - 2\eta}{1 + \eta},
\end{equation}

At loads below $\sigma_\mathrm{HEL}$ the elastic regime is realized: elastic waves propagate with the speed $c_l$. When $\sigma_\mathrm{HEL}$ is exceeded, waves propagate in the splitting regime: the elastic wave propagates faster than the bulk one and separates from it. However, at such loads the longitudinal sound speed $c_l$ reaches the maximum which corresponds to $\sigma_\mathrm{HEL}$. The bulk sound $c_b$ speed is not limited, and the overdriven regime is realized for the most intense loads: the speed of an elastic wave is exceeded by the speed of a plastic one, so that only one wave is observed.

\subsection{Smoothed Particle Hydrodynamics}

\begin{figure}[t]
\includegraphics[width=1\linewidth]{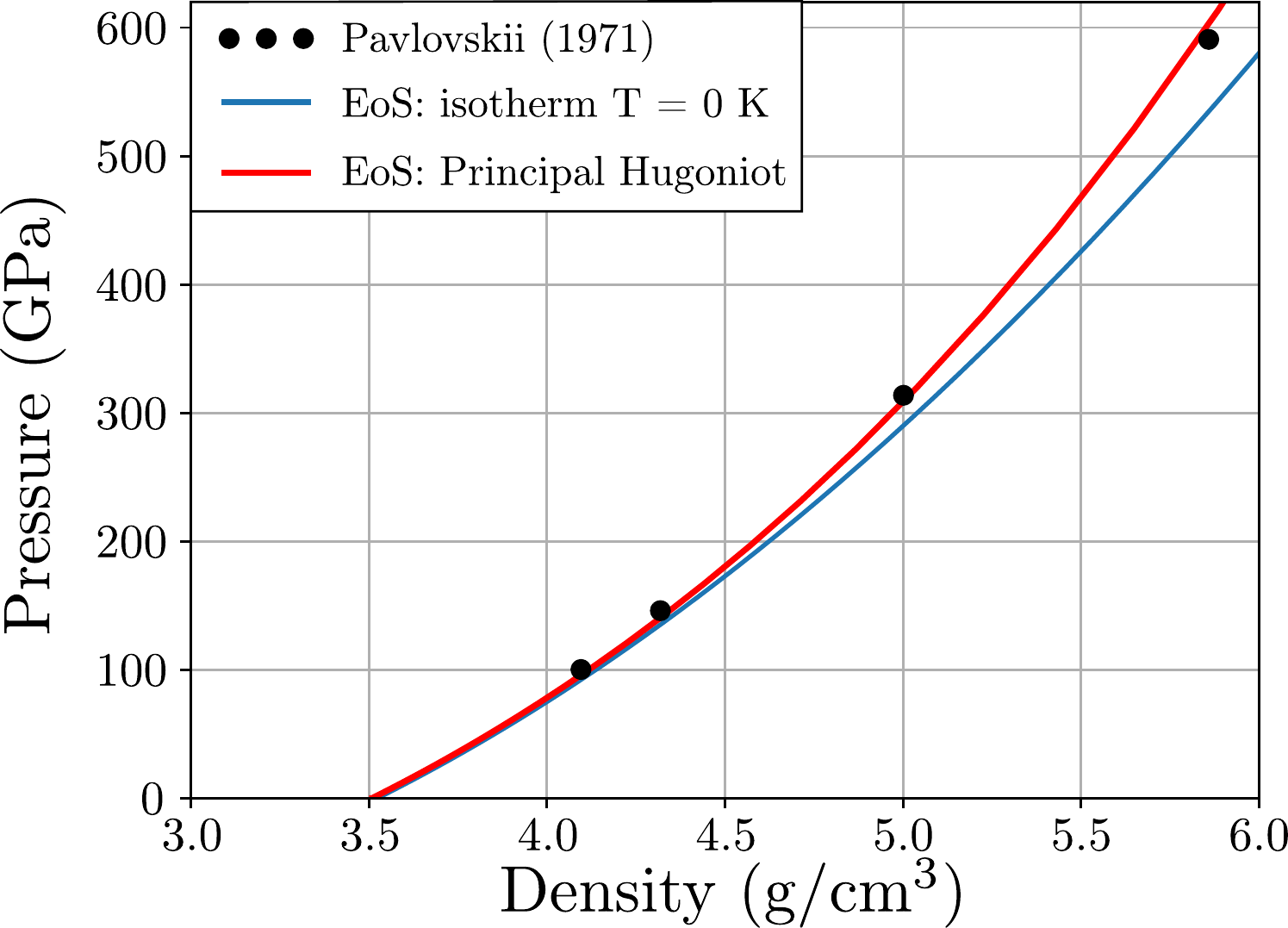}
\caption {\label{fig:diamond-eos} The principal Hugoniot and zero isotherm for diamond according the EoS~\cite{Lomonosov:1994} and the data~\cite{1971:FTT:Pavlovskii}.}
\end{figure}

\begin{figure*}[t]
\includegraphics[width=1\linewidth]{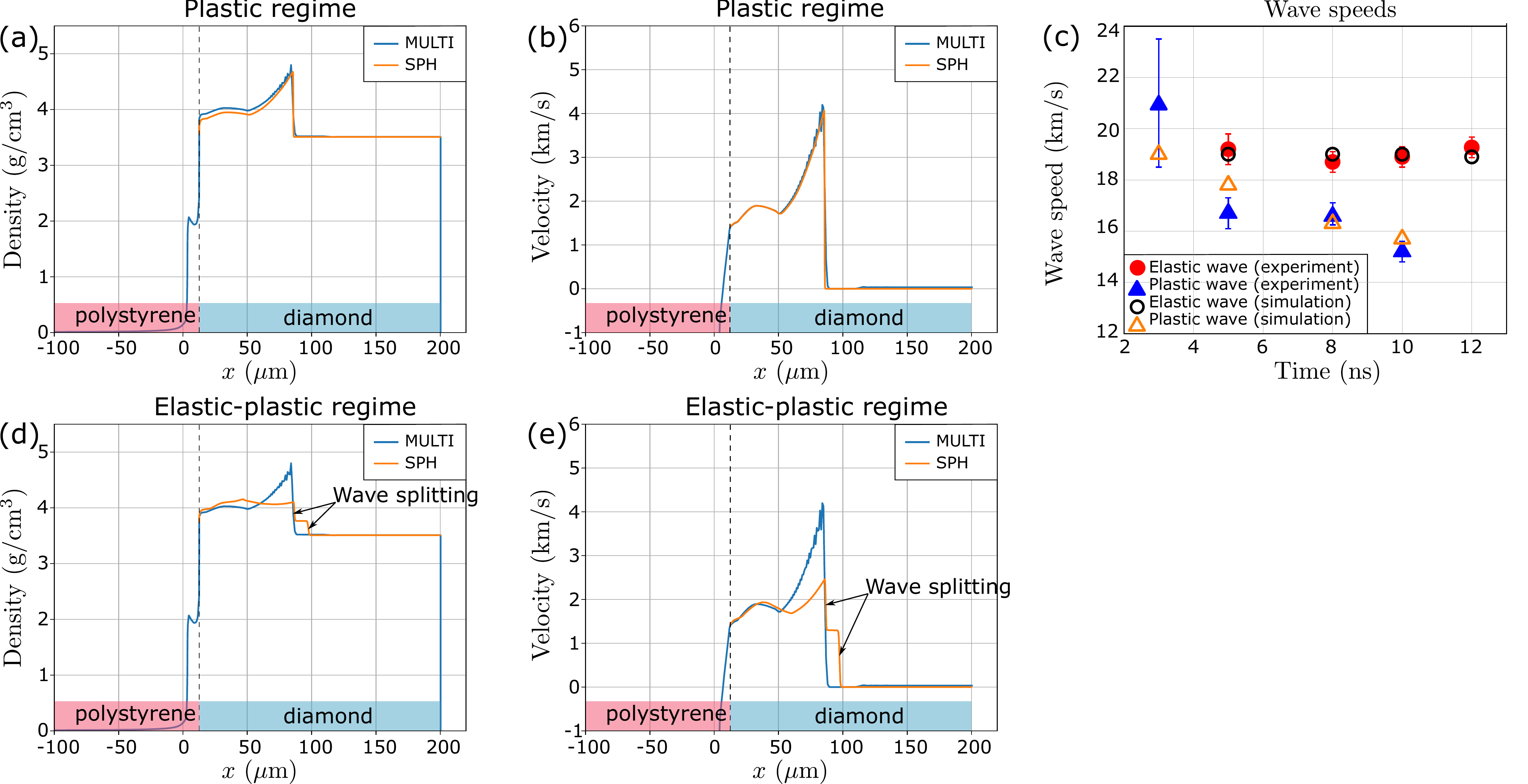}
\caption {\label{fig:piston-MULTI-vs-SPH} (a), (b) Validation of the boundary condition extracted from MULTI results in SPH with pure plastic model of diamond. (c) Comparison of wave speeds obtained in the experiment and the one-dimensional simulation. (d), (e) Elastic and plastic wave splitting observed in SPH with the same boundary condition in the diamond with strength $Y = 70\un{GPa}$.}
\end{figure*}

Smoothed particle hydrodynamics (SPH) method~\cite{2002:JCP:Parshikov} provides the following discretization scheme for Eqs.~\eqref{continuity}--\eqref{energy}:
\begin{equation}
 \label{continuity_SPH}
 \frac{ d \theta_i}{dt} = 2\sum_{j}\frac{ m_j }{\rho_j}
 \left(\mathbf{U}^*_{ij} - \mathbf{U}_i\right)\cdot \nabla W(|\mathbf{r}_i - \mathbf{r}_j|, h).
\end{equation}
Momentum conservation~\eqref{momentum} in our SPH implementation is guided by
\begin{equation}
 \label{momentum_SPH}
   \frac{ d\mathbf{U}_i}{dt}= -\frac{2}{\rho_i}\sum_{j}\frac{m_j}{\rho_j} \bm{\upsigma}_{ij}^* \nabla W(|\mathbf{r}_i - \mathbf{r}_j|, h),
 \end{equation}
and the energy conservation~\eqref{energy} equation is
\begin{equation}
\label{energy_SPH}
\frac{ d E_i}{dt}=
-\frac{2}{\rho_i}\sum_{j}\frac{m_j}{\rho_j} \bm{\upsigma}^*_{ij} \mathbf{U}^*_{ij} \cdot \nabla W_{ij}(|\mathbf{r}_i - \mathbf{r}_j|, h).
\end{equation}
Here $m_i$ is the particle mass, $\rho_i$ is the particle density, $U_{ij}^*$ is the velocity of contact surface between particles, $\bm{\upsigma}^*_{ij}$ is the stress tensor at the contact between particles, $W(r,h)$ is the smoothing kernel function. Similarly, $\dot{\bm{\upvarepsilon}}$ and $\mathbf{\Omega}$ can be evaluated by discretizing Eqs.~\eqref{eq:strain-rate} and \eqref{eq:angular-velocity} using which the corresponding stress deviator tensor $\hat{\mathbf{S}}$ is obtained via Eq.~\eqref{eq:stress-tensor}. The latter is corrected according the criterion~\eqref{eq:yield-criterion}:
\begin{equation}
\bm{S} = \frac{\upsigma_e}{Y}\hat{\mathbf{S}}.
\end{equation}

\subsection{Equation of State}

The equation of state (EoS) for diamond used in SPH simulations is constructed using the generalized Mie--Gruneisen form where the Gruneisen parameter $\Gamma$ depends on the material density~\cite{Lomonosov:1994}. The cold (reference) energy in the region of compression $x_c = \rho / \rho_{0c} > 1$, where $\rho_{0c}$ is the density at $P = 0$ nad $T = 0$, is described by:
\begin{equation}\label{eq:eos-e-ges}
e_{c}(\rho) = \frac{3}{\rho_{0c}}\sum_{i = 1}^{N_{eos}}\frac{a_i}{i}\left(x_c^{i/3} - 1\right),
\end{equation}
where $N_{eos}$ is the number of terms (2, 5 or 9). They are defined in a way that $e_c(\rho_{0c}) = 0$, and should agree with the experimental data and first principles modelling at high pressures. At rarefaction ($x_c < 1$) the specific energy is given by a polynomial:
\begin{equation}
    e_{c}(\rho) = \frac{B_{0c}}{\rho_{0c}(m - n)}\left(\frac{x^m}{m} - \frac{x^n}{n}\right) + E_{sub},
\end{equation}
which coefficients are defined to agree the reference data on the cohesion energy $e_c(\rho \rightarrow 0) = E_{sub}$ and the condition $P_c(\rho_{0c}) = 0$.

The Gruneisen parameter is:
\begin{equation}
\label{eq:eos-gamma-ges}
\Gamma(\rho, e) = \gamma_i + \frac{\gamma_c(\rho) - \gamma_i}{1 + x^{-2/3}[e - e_c(\rho)]/e_a},
\end{equation}
where $\gamma_c(\rho)$ is for the low temperatures, while $\gamma_i$ is used for the high temperature plasmas. The energy of anharmonicity $e_a$ corresponds to the transition energy between asymptotic expressions for high temperatures. 

The reference data on diamond may be found in the shock wave database~\cite{rusbank}, where the equation of state for used in our simulations diamond corresponds KEOS5. Figure~\ref{fig:diamond-eos} demonstrates the zero isotherm and the principal Hugoniot for diamond according this EoS.

\subsection{One dimensional simulations: adjusting HEL and shear modulus}

\begin{figure}[t]
\includegraphics[width=1\linewidth]{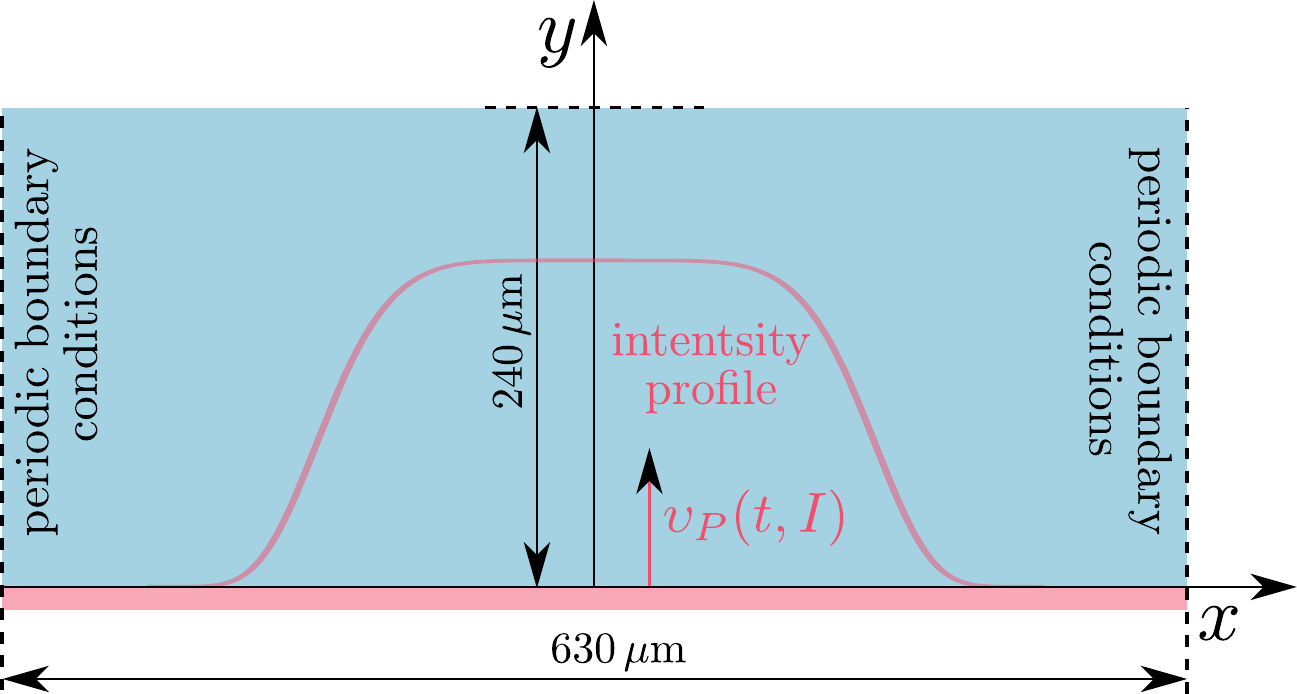}
\caption {\label{fig:SPH-2D-setup} Problem setup for the two-dimensional SPH simulation of the diamond loading with an intense laser pulse. The diamond sample is a rectangle with sizes $630\un{\mu m}\times 240\un{\mu m}$ placed in periodic boundary conditions along $x$-axis. The boundary at $y = 0$ moves with the velocity $\upsilon_P(t, I)$ which is the interpolation of several MULTI profiles given in Fig.~\ref{fig:multi-piston-time}. }
\end{figure}

\begin{figure}[t]
\includegraphics[width=0.95\linewidth]{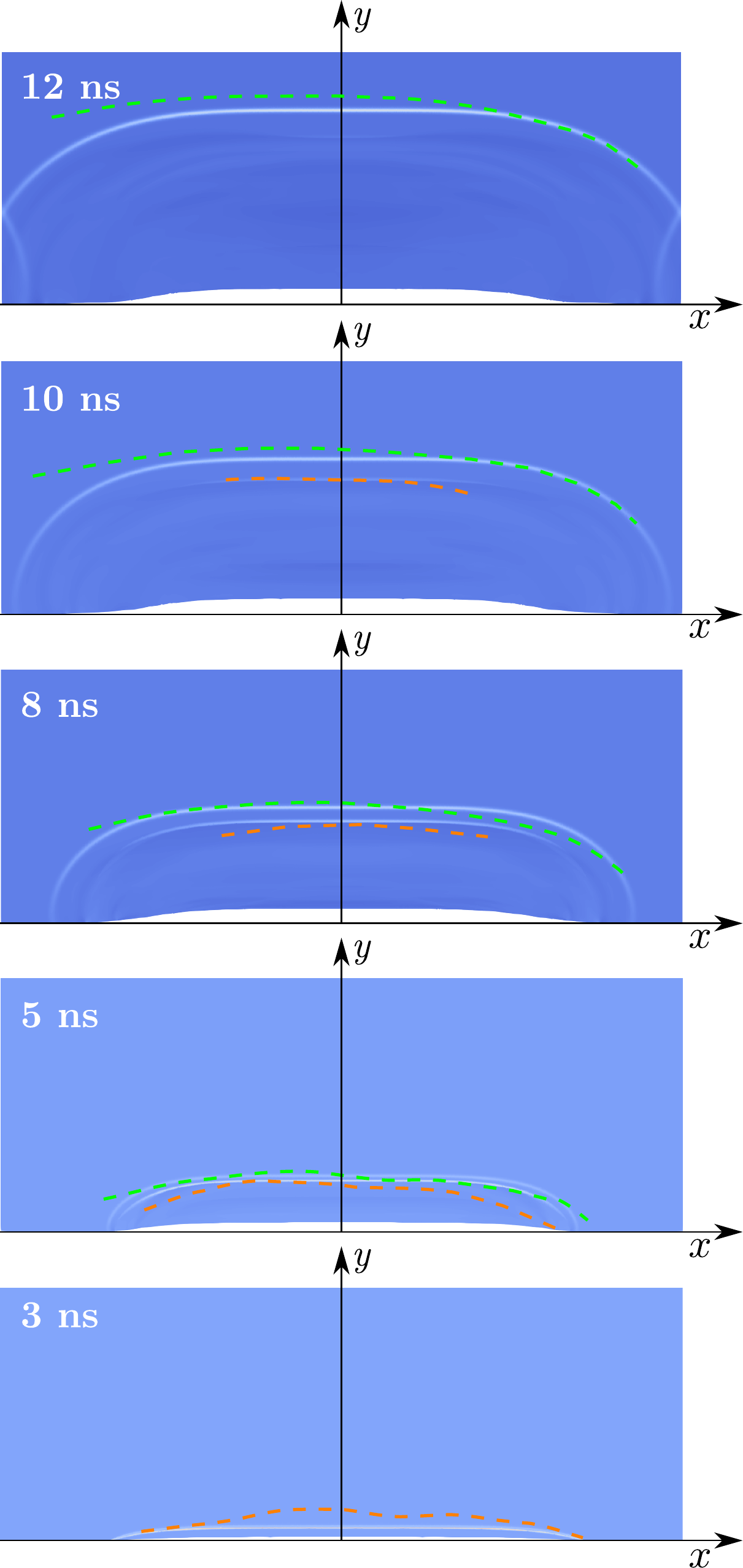}
\caption {\label{fig:SPH-2D-results} The two-dimensional maps for the strain rate are used to enhance visualization of the shock fronts. The experimental shock fronts were digitized from Fig.~\ref{fig:wave-propagation-exp} and placed as dashed curves at the corresponding SPH results.}
\end{figure}

\begin{figure*}[t]
\includegraphics[width=0.85\textwidth]{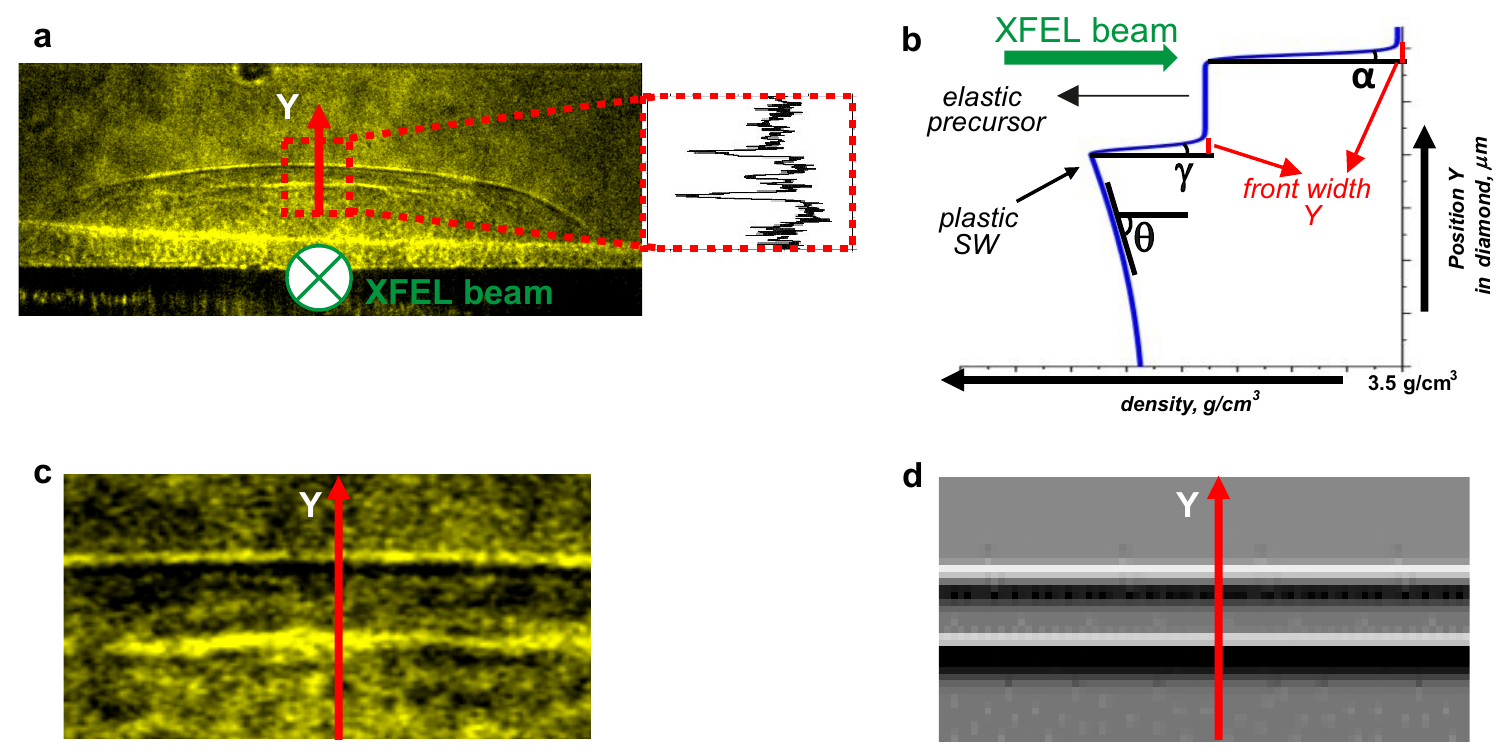}
\caption {\label{wpg_1} Estimation of the shock wave front widths using the WavePropogator simulation: (a) Phase-contrast LiF image of shock waves in diamond for a time interval of 8 ns. Shown is the direction of propagation of the SACLA XFEL probing beam and the intensity distribution within the area selected for analysis. (b) Schematic of the target density setting for the simulation with marked parameters considered in the calculations. (c) Zoomed frame within the red box in image (a). (d) Calculated phase-contrast pattern corresponding to the experimental area in image (c).}
\end{figure*}

\begin{table*}
\caption{\label{table:material_props} Results of SPH simulation for monocrystalline diamond $<$100$>$ (Hugoniot elastic limit value $Y = 0.7\un{Mbar}$; Poisson coefficient = 0.15; Bulk modulus = 4.81 Mbar; Shear modulus = 4.39 Mbar). 
1-elastic precursor and 2-plastic SW: mean particle velocity $u_1$ and $u_2$, diamond density $\rho_1$ and $\rho_2$, pressure $P_1$ and $P_2$.}
\label{tab:a}
\tabcolsep7pt
\begin{tabular}{ccccccccc}
\hline
Time, ns & $u_1$, km/s & $V_1$, km/s & $\rho_1$, g/c$m^3$ & $P_1$, Mbar & $u_2$, km/s & $V_2$, km/s & $\rho_2$, g/c$m^3$ & $P_2$, Mbar  \\
\hline
       3  &   -    &  -   &  -  &   -  &   5.32   &    19     &  4.91  &   3 \\
       5  &   1.3   &  19   &  3.76  &    0.4  &   3.42  &    17.2   &  4.38  &   1.62 \\
       8  &   1.3   &  19   &  3.76  &    0.4  &   2.12  &    16.3   &  4  &    0.83 \\
      10  &   1.3   &  19   &  3.76  &    0.4  &   1.6  &    15.7   &  3.85  &    0.55 \\
      12  &   1.3   &  19   &  3.76  &    0.4  &   -  &     -     &  -  &    - \\
\hline
\end{tabular}
\tabcolsep7pt
\end{table*}

The main reason we applied SPH for the considered problem is its ability to model the intense loads of materials with strength. Diamond has uniquely high HEL of $50$--$80\un{GPa}$ and is subjected to loads up to $400\un{GPa}$ in our experiments. However, our SPH implementation is lacking the radiation transport support, so that MULTI predictions of the polystyrene-diamond interface are performed to setup the appropriate boundary condition. Fig.~\ref{fig:MULTI} shows the interface velocity profile obtained in MULTI at laser pulse intensity of $I = 6 \times 10^{12}\un{J/cm^2}$ which is interpolated for the further use in SPH. One may notice, that the initial velocity jump exceeds $5\un{km/s}$ and is followed by the fast unloading after $1\un{ns}$ to about $2\un{km/s}$. The latter amplitude is supported by the plasma pressure in corona for about $3\un{ns}$, after which follows the gradual release to zero pressure. Similar response is observed for the density profile, which is also interpolated to evaluate the corresponding pressure in diamond at the interface with polystyrene.

To validate the diamond response in SPH with the boundary condition based on MULTI results we performed simulations with pure plastic model ($Y = 0$). The diamond sample of $200\un{nm}$ length in one-dimensional SPH consists of $4000$ particles. The boundary condition is made of $10$ diamond particles with the predefined density and velocity according to MULTI data (Fig.~\ref{fig:multi-interp}): these particles transfer the laser induced pulse to the bulk of the main diamond sample. The resulted profiles after $5\un{ns}$ of wave propagation are shown in Fig.~\ref{fig:piston-MULTI-vs-SPH}a, \ref{fig:piston-MULTI-vs-SPH}b. One may notice a very good agreement of the density and velocity profiles obtained in MULTI with the realistic model of polystyrene ablator and in SPH with the appropriate boundary condition at the interface. Both the amplitude and the wave speeds agree well which indicates the validity of the applied boundary condition.

Next, the model of diamond with the constant shear strength $Y = 70\un{GPa}$ is considered. The corresponding density and velocity profiles are shown in Fig.~\ref{fig:piston-MULTI-vs-SPH}d, \ref{fig:piston-MULTI-vs-SPH}e. One may notice the appearance of the elastic and plastic waves splitting, which is observed in our experimental data with similar loading. Having the measured distances between the wave fronts and their propagation speeds we can adjust the diamond $\upsigma_{HEL}$ and the Poisson coefficient $\eta$ (or the related shear modulus $G$) in our model for the best agreement with experiments at various pulse intensities. Figure~\ref{fig:piston-MULTI-vs-SPH}c compares the wave speeds obtained in the experiments and our simulations. It appears that $\upsigma_{HEL} \simeq Y = 70\un{GPa}$ and $\eta = 0.15$ provide the best fit for the experimental data. Table 1 shows the parameters of the diamond medium in the region of the observed shock waves along the direction $Y$.

\subsection{Multi-dimensional simulations}

Simulation setup for two-dimensional simulations is given in Figure~\ref{fig:SPH-2D-setup}. A rectangular sample of $630\un{\mu m}$ along $x$-axis and $240\un{\mu m}$ along $y$-axis is represented with SPH-particles of size $D = 0.25\un{\mu m}$. The quite fine mesh is necessary to resolve the wave structure at the beginning of loading and further separation of the elastic precursor from the plastic wave. Periodic boundary conditions are applied along $x$-axis, while $y$ direction is subjected to loading via the ablator response model given in Fig.~\ref{fig:multi-piston-time}.

The result of the SPH simulation of diamond loading with the laser pulse of $I = 6 \times 10^{12}$ is given in Fig.~\ref{fig:SPH-2D-results}. For convenience, the experimental shock fronts were digitized and placed as dashed curves at the corresponding SPH results. The two-dimensional maps for the strain rate~\eqref{continuity} are used to enhance visualization of the shock fronts. One may notice, that the elastic and plastic waves are not separated at $3\un{ns}$ which is observed in the simulation and the experiment, but the latter wave have propagated further. This may happen due to non-uniform heating of the ablator which results in the wave front distortion. At $5\un{ns}$ the separation of the elastic precursor occurs both in the experiment and simulation, and the observed and predicted wave speeds become close. At $8\un{ns}$ waves are separated further, and the positions of simulated shock fronts agree well with the experiment. The rarefaction wave overtake the plastic wave at about $10\un{ns}$, so that it almost disappear: the predicted position of the remaining part still agree with the experiment. Finally, the plastic wave disappeared at $12\un{ns}$, but the elastic one propagates further. There is also quite good agreement between the simulation and the experiment: the initial distortion of the front affects the position of elastic precursor at later times which is not taken into account in SPH.

The presented SPH simulation demonstrates the ability to reproduce the complex phenomena of waves splitting in the bulk of material sample. The series of experimental images can be used to adjust the strength model precisely by tracking positions of the shock fronts.

\section{Estimation of density gradient in Shock Waves}

As part of the study of shock wave morphology, we estimate a density gradient (shock wave width) using a method based on the analysis of monochromatic X-ray images with high spatial resolution in phase contrast and modelling of shock propagation by the Code WavePropagator. Images corresponding to times $t = 3\un{ns}$ (one SW is observed) and $t = 8\un{ns}$ (two SWs are clearly visible) were selected for analysis. On the 2D phase contrast image of LiF, the region where the shock wave front can be considered approximately flat was chosen to minimise the blurring effect throughout the diamond volume (red region in Fig.~\ref{wpg_1}a). A number of independent parameters were introduced to simulate the propagation of the probe beam through the shocked diamond sample. These include the density of the compressed material, the density gradient of the shock wave and the width of the shock wave plate. The method was applied to determine the density gradients of the shock waves observed in the diamond sample at a delay of 3 ns (one shock wave) and 8 ns (two shock waves) between the pump and probe pulses. In our case, the following parameters were used, Fig.~\ref{wpg_1}b):

\begin{figure}[t]
\includegraphics[width=1\linewidth]{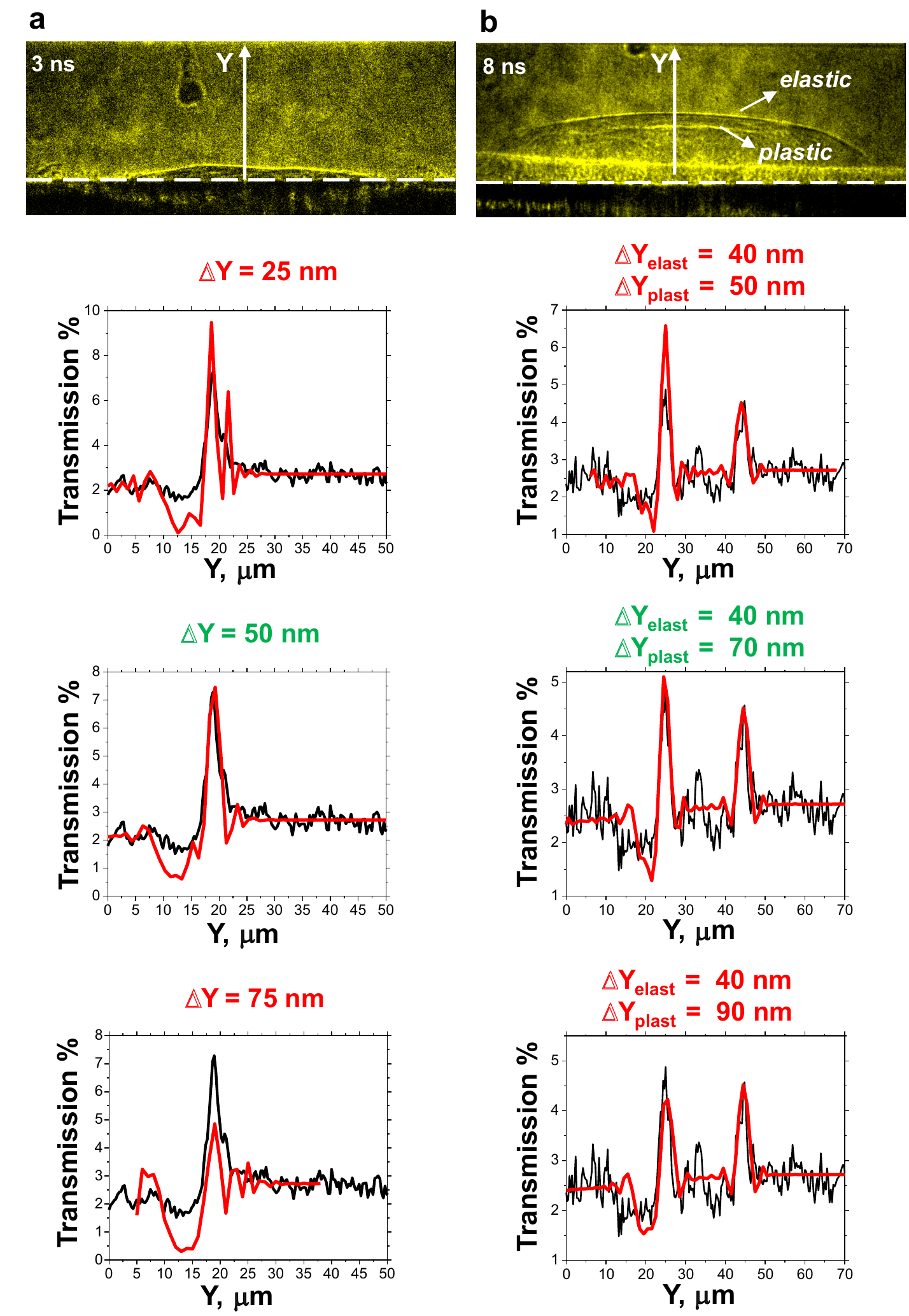}
\caption {\label{wpg2} { Results of the estimation of the front width $\Delta Y$ of shock waves observed in diamonds} at times $t = 3\un{ns}$ (a) and $t = 8\un{ns}$ (b). The left PC images show the areas where the intensity distribution was recorded for the experimental profile.}
\end{figure}

\begin{itemize}
\item angle $\alpha$ associated with density gradient of front of elastic SW; 
\item angles $\gamma$, $\theta$ associated with density gradients of front and rear slope of plastic SW; 
\item density amplitudes for elastic and plastic SW.
\end{itemize}

By fixing 3 parameters (the angle $\theta$ and density amplitudes for elastic and plastic SW, derived from the results of the SPH code simulations) and by varying the angles $\alpha$ (corresponding to the SW width), we found the particular solutions that describe the experimental PCI profile. As an example, Fig.~\ref{wpg_1}(c,d) shows the experimental and calculated patterns for time $t = 8\un{ns}$.

At first, we consider the time at which the structure of the pair waves is not yet observed ($t = 3\un{ns}$). Fig.~\ref{wpg2}a shows a comparison of the experimental and calculated intensity distributions in the plane of the LiF detector when the SW front width is set in the range $Y = 25$--$75\un{nm}$. The best agreement between the data is observed for the width $Y = 50\un{nm}$. For the time of $8\un{ns}$, the front width of the elastic precursor $Y_\mathrm{elast} = 40\un{nm}$ and that of the slower plastic SW was found to be $Y_\mathrm{plast} = 70\un{nm}$, Fig.~\ref{wpg2}b. These estimates are an order of magnitude higher than the values expected by theory (of the order of the interatomic lattice spacing), which could be due to two reasons: 1) we do not observe a flat front of the shock waves, but a curved one, which blurs the experimental profile of the phase contrast image; 2) the detector resolution obviously increases the experimentally measured profile, which is then compared to the model.

\section{Conclusion}
The achieved excellent agreement between the experimental data and continuum mechanics modelling not only paves the way for direct measurement of the dynamic yield strength of materials as a function of strain rate, but also highlights the usefulness of these facilities for the study of high-speed crack dynamics and unusual stress-induced solid-state phase transitions. These transitions could have a significant impact on the development of new materials in industry and enable the invention of interesting mechanical devices. The presented experimental approach to the study of shock waves in solids opens up new possibilities for the verification and construction of the equations of state of matter in the ultra-high pressure regime, which are relevant for the solution of a variety of problems in high energy density physics.

\section*{References}
\bibliography{library}

\end{document}